\documentclass[journal]{IEEEtran}

\usepackage[pdftex]{graphicx}

\usepackage{amsmath}
\usepackage{amssymb}

\usepackage{cite}

\usepackage{xcolor}

\newcommand{\const}{\mathop{\rm const}\nolimits}

\begin{document}

\title{Randomly spaced phase-only transmission combs for femtosecond pulse shaping}


\author{Konstantin B. Yushkov and Vladimir Ya. Molchanov

\thanks{This work was supported in part by the Ministry of Science and Higher Education of the Russian Federation (Gov. Decree No. 211 of 16 March 2013), National University of Science and Technology ``MISIS'' (K2-2020-007), and by the Russian Foundation for Basic Research (Project 18-29-20019).}
\thanks{K. B. Yushkov, and V. Ya. Molchanov, are with the National University of Science and Technology ``MISIS'', 4 Leninsky Prospekt, Moscow 119049, Russia (e-mail: konstantin.yushkov@misis.ru)}
\thanks{Manuscript compiled \today.}}


\maketitle

\begin{abstract}
We present a new Randomized Multiple Independent Comb Shaping (RandoMICS) algorithm based on phase-only tailored transmission for ultrashort laser pulse replication. The benefit of this method is satellite-free generation of programmable laser pulse sequences. The result is achieved by creating a transmission function as a stochastic comb of disjoint segments of optical frequency continuum with numerically optimized segment width distribution. The algorithm is realized by generating a regular aperiodic comb and random permutations of its elements. Experimental demonstration is performed with an acousto-optic pulse shaper providing broadband multi-window transmission function with arbitrarily variable widths of the segments. Suppression of undesired satellite pulses by the factor of 8 is demonstrated as well as generating pulse replicas with extended usable delay range compared to phase-only pulse shaping with periodic transmission combs.
\end{abstract}


\section{Introduction}

The concept of adaptively reconfigurable resolution in optics and photonics has been recently discovered as a new paradigm to improve performance of different signal processing techniques from microwave filters to hyperspectral imaging~\cite{JiangYanMarpaung18,CongYamamotoInoue19,YushkovMakarovMolchanov19}. Applications of this approach are fueled by new architectures of photonic devices and adaptive algorithms for controlling them. Similar techniques of tailoring optical fields found broad coverage in optics, photonics, and laser technology. For example, effective computational optimization algorithms facilitated design of aperiodic structures with suppressed constructive interference that has been used in laser pulse shaping, inertial confinement fusion, and diffraction grating technology~\cite{RoccaOliveriMassa11,AlbrightYinAfeyan14,TorcalSanchez16,Vishnyakov18,IkonnikovArkhipkinVyunishev19}.

Acousto-optic (AO) diffraction can be used for programmable laser pulse shaping, providing both phase and amplitude modulation of light~\cite{DuganTullWarren97,VerluiseLaudeChengEtal00,GaoHerriotWagner06,NagChaphekarGoswami10,MolchanovYushkov14,YushkovMolchanovOvchinnikovChefonov17}.  Most of the achievements in this field are based on availability of digital radio frequency (RF) arbitrary waveform generators (AWGs) that are capable of synthesizing precision spread-spectrum signals for feeding AO devices. Despite a broad scope of techniques and applications, the potential of flexible reconfiguration of diffraction parameters in AO devices has not yet been exploited in full.

\begin{figure}[t]
  \centering
  \includegraphics[width=\columnwidth]{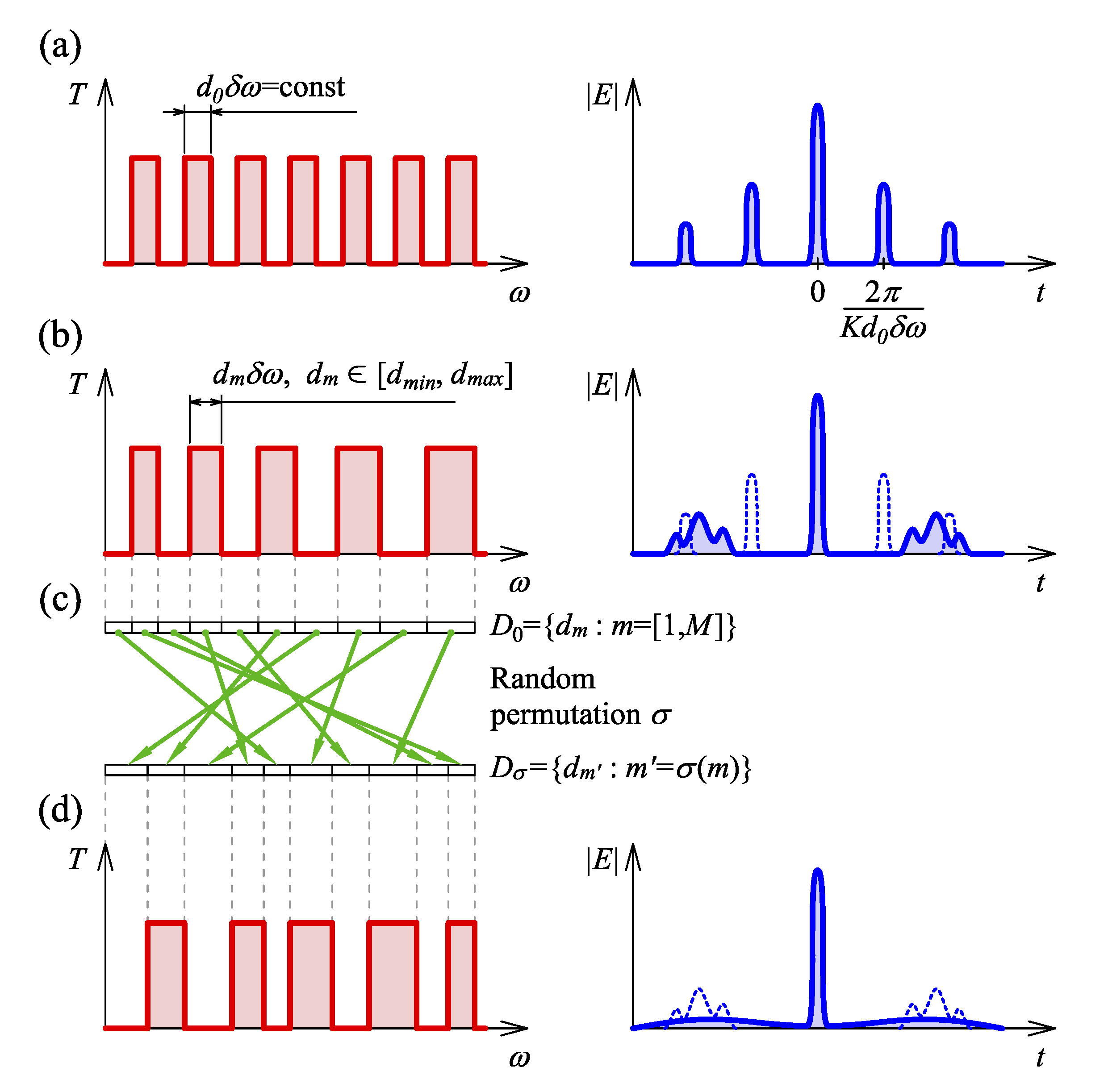}\\
  \caption{The principle of arbitrary-spaced comb: (a) an equally spaced comb produces satellite pulses with deterministic positions at $t_p=2\pi p/(K d_0\delta\omega)$; (b) an aperiodic comb produces a broadened satellite with lower amplitude; (c) a permutation $\sigma$ is used to shuffle an aperiodic comb; (d) a randomly spaced comb produces a broad but low-amplitude pedestal. Left: spectral transmission $T(\omega)$; right: electrical field amplitude $E(t)$.}\label{fig1:combs}
\end{figure}

In this work we propose and experimentally demonstrate the principle of reconfigurable resolution for suppression of satellite pulses in phase-only ultrashort laser pulse replication. Replication of laser pulses is a special type of pulse shaping, which has been previously demonstrated both with AO pulse shapers and with Fourier transform (FT) pulse shapers based on spatial light modulators (SLMs)~\cite{WeinerHeritageKirschner88,WeinerLeaird90,WeinerLeairdPatelWullert92,WefersNelson95,DuganTullWarren97,DorrerSalin98,VerluiseLaudeChengEtal00,RundquistEfimovReitze02,PestovLozovoyDantus09,YushkovMolchanovOvchinnikovChefonov17}.
Combination of a pulse shaper with the Michelson interferometer has been demonstrated for generation of long pulse trains~\cite{BitterMilner16}. Some relevant applications of programmable ultrashort pulse trains are self-referenced pulse metrology~\cite{GallerFeurer08,MohringBuckupMotzkus10,KearnsMehlenbacherJonesZanni17,Dorrer19}, pump-probe experiments~\cite{Jonas03,KuhsLutherKrummel19}, generation of THz radiation~\cite{AhnEfimovAverittTaylor03,OvchinnikovMolchanovYushkov16_eng}, \emph{etc}.

Significant research efforts have been made to develop phase-only laser pulse shaping methods mainly for application with FT SLM  shapers~\cite{WeinerLeaird90,BaumertBrixnerSeyfried97,RundquistEfimovReitze02,WilsonSchlupBartels07,PestovLozovoyDantus09,GalvanPortillaHernandez13,FarfanEpsteinTurner18}. Iterative feedback-controlled computation of the SLM transmission pattern is a commonly used option in pulse shaping. Baumert \emph{et al.} developed evolutionary algorithm for adaptive phase-only pulse shaping~\cite{BaumertBrixnerSeyfried97}. Rundquist \emph{et al.} applied the Gerchberg-Saxton algorithm~\cite{RundquistEfimovReitze02}. Since then, different iterative algorithms including those involving machine learning have been developed and demonstrated improved  convergence speed and efficiency~\cite{NagChaphekarGoswami10,GalvanPortillaHernandez13,FarfanEpsteinTurner18}. An alternative is to use straightforward algorithms for explicit calculation of the shaper transmission. Such algorithms are inherently faster and less resource-demanding than the iterative ones. Within this approach, Wilson \emph{et al.} demonstrated that a high-resolution phase-only SLM can simulate phase-and-amplitude spectral modulation by means of high-frequency phase modulation~\cite{WilsonSchlupBartels07}. Weiner \emph{et al.} pioneered research on the problem of pulse train generation and represented the phase-only transmission function as several interleaved subsets of the whole frequency interval~\cite{WeinerHeritageKirschner88,WeinerLeaird90}. This approach was further developed by Pestov \emph{et al.} as the Multiple Independent Comb Shaping (MICS)~\cite{PestovLozovoyDantus09}. Hereinafter, we follow this paradigm and treat a ``comb'' as a union of disjoint segments of the optical frequency continuum (the ``comb teeth'') .

A periodic wavelength comb inevitably produces undesired satellite laser pulses~\cite{PestovLozovoyDantus09,VaughanFeurerStoneNelson06,YushkovMolchanovOvchinnikovChefonov17}. The satellites form a regular structure with the interval inversely proportional to the comb period $K d_0\delta\omega$, where $K$ is the number of replicas (duty cycle of the comb), $\delta\omega$ is the minimum resolvable element, and $d_0$ is the binning factor. Interference between the replicas and the satellite pulses limits the maximum replica delay $\tau$ as one half of the first satellite position, $\max \tau = \pi/(K d_0\delta\omega)$. The negative effect of the satellite pulses increases with the number of generated replicas $K$~\cite{PestovLozovoyDantus09}. That implies a higher requirement on the shaper resolution to obtain larger delays. In liquid crystal SLMs the comb period is fixed and determined by the pixel pitch, and the only way to change it is to use pixel binning. On the contrary, AO pulse shapers can operate with a continuously variable software-defined comb tooth width~\cite{YushkovMolchanovOvchinnikovChefonov17}.

We develop and elaborate the principle of MICS by means of generating uneven teeth of the comb either in a regular way or randomly. Further we demonstrate that using an aperiodic comb generated with a Randomized Multiple Independent Comb Shaping (RandoMICS) algorithm results in smoothing out and suppression of the satellite pulses. The principle of creating a random optical comb is illustrated in Fig.~\ref{fig1:combs}. The algorithm is realized by applying random permutations to a regular aperiodic comb of frequency segments. The width of each segment is assumed to have the lower and upper bounds $d_{min}\delta\omega$ and $d_{max}\delta\omega$. The effect of segment width distribution is analyzed and constrained optimization problem is solved using differential evolution (DE) genetic algorithm~\cite{Coello02}. Performance of the RandoMICS algorithm is assessed experimentally.

\section{Acousto-optic pulse shaper}\label{sec-AO}
Two common schemes of laser pulse shaping are the FT SLM in spectral domain and direct time-domain pulse shaping using AO diffraction in a collinear or close-to-collinear configuration~\cite{Weiner11}. AO pulse shapers have a smaller footprint than FT systems and support the laser pulse repetition rate up to 100 kHz. The update rate of liquid crystal SLMs is below 1 kHz. On the other hand, liquid crystal SLMs can continuously hold the transmission function unlike AO devices, since the latter rely on diffraction of light by traveling nonstationary ultrasonic wave packets. Another aspect of comparing is complexity of transmission functions. An AO pulse shaper can simultaneously perform phase and amplitude modulation of the spectrum, while a single SLM is capable of either phase-only or amplitude-only modulation. Keeping in mind the latter restriction, we used an AO pulse shaper in experiments to demonstrate and quantitatively measure the effect of satellite suppression in an irregular grid apart from pulse replication.

AO filters are the devices with electronically adjustable transmission function. The simplest methods to achieve variable transmission bandwidth and multiple transmission windows are linear phase modulation and mixing several single-frequency RF signals feeding the filter~\cite{Magdich80_eng,MagdichMolchanovPonomareva84_eng,ShnitserAgurok97}. Precise transmission function tailoring is based on digital methods of RF signal synthesis, namely fast Fourier transform (FFT) accompanied by proper mapping from optical to RF domain~\cite{MolchanovYushkov14,SPIE19_11210}. In this case, time and frequency are discrete variables, hence the width of transmission windows can not be changed continuously. However, the typical sampling grids both in time and in frequency domains are dense enough to provide quasi-continuous comb tooth width changes. For this reason, we focused on programmable AOTFs to implement RandoMICS pulse shaping.

Direct time-domain pulse shaping was performed in quasicollinear AO filter geometry~\cite{MolchanovVoloshinovMakarov09_eng,MolchanovEtal09}. The high-definition AO dispersion delay line (AODDL) was designed and fabricated in-house~\cite{DidenkoEtal15_eng}. The configuration of the AODDL was optimized for broadband Ti:sapphire laser emission. The 80-mm-long TeO$_2$ crystal had the acoustic time aperture of $t_a=102.4$~$\mu$s. The instantaneous processed bandwidth of laser radiation was 150 nm centered at 795 nm that corresponded to the phase-matched ultrasonic bandwidth of $\Delta F =17.05$~MHz. The number of frequency samples, $N= 2 \Delta F t_a =3494$, was limited by the Whittaker-Kotelnikov-Shannon sampling theorem~\cite{SPIE19_11210}. The resolution passband of the AODDL was 0.24~nm corresponding to the minimum binning factor $d_{min}=5$.

\section{Random spectral combs}\label{sec-combs}
\subsection{Definition}
\begin{figure}[t]
  \centering
  \includegraphics[width=\columnwidth]{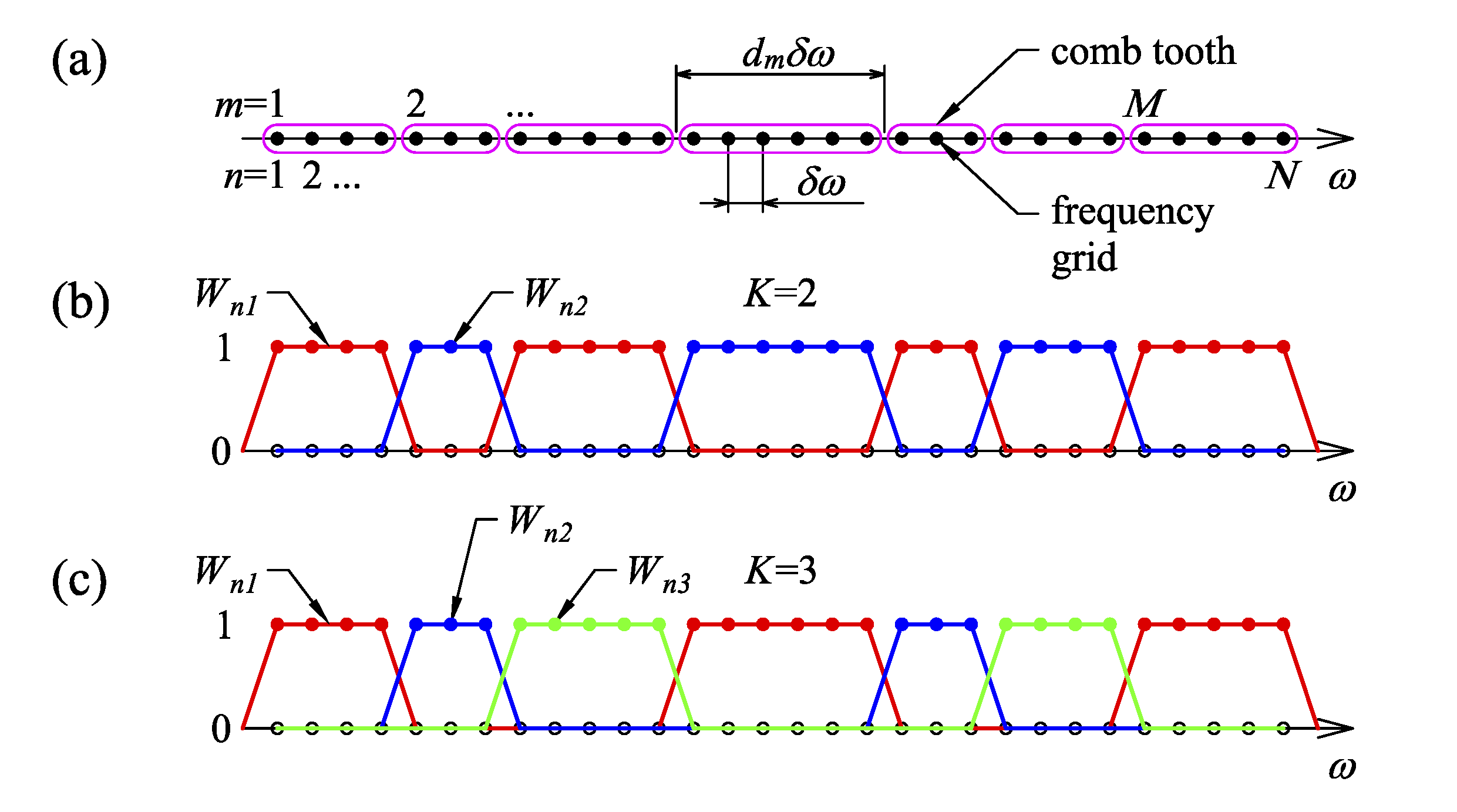}\\
  \caption{Formation of complementary interleaved aperiodic transmission windows corresponding to different number of replicas $K$: (a) the comb is obtained by uneven binning of $N$ frequency grid points into $M$ comb teeth; (b,c) different sets of orthogonal window functions $W_{nk}$ are obtained from the same comb, $k=[1,K]_{\mathbb{N}}$ is the running index of the replica.}\label{fig2:grid}
\end{figure}

Hereinafter we use an equally-spaced grid of optical frequencies $\{\omega_n:n=[1,N]_{\mathbb{N}}\}$ with the increment $\delta\omega$. Any comb defined on this grid has a disambiguous representation by an integer-valued vector of tooth widths
\begin{equation}\label{eq-D}
D_0=\{d_m:m=[1,M]_{\mathbb{N}},d_m\in\mathbb{N}\},
\end{equation}
where $M$ is the total number of the comb teeth. Each $d_m$ is considered as an integer binning factor varying from tooth to tooth. This comb vector is used to define the transmission as a grid function, $T=\{T_n:n=[1,N]_{\mathbb{N}}\}$. In the case of generating $K$ replicas of an ultrashort pulse having the delays $\tau_k$ and amplitudes $A_k$, the transmission is defined as
\begin{equation}\label{eq-Tn}
  T_{n} = \sum_{k=1}^{K} A_k \exp (i\omega_n \tau_k) W_{nk}
\end{equation}
where $W_{nk}$ are orthogonal discrete window functions
\begin{equation}\label{eq-Djk}
W_{nk}=
\left\{
\begin{aligned}
&1,\qquad\mbox{if}\quad (m'(n)-k)\,{\rm mod}\,K=0;\\
&0, \qquad\mbox{otherwise};\\
\end{aligned}
\right.
\end{equation}
$x\,{\rm mod}\,y$ is the modulo operator (remainder), and $m'(n)$ is the index of the comb tooth that contains the frequency $\omega_n$:
\begin{equation}\label{eq-mprime}
m'(n) = \min \left\{m : n \leqslant  \sum_{j=1}^{m} d_j\right\}.
\end{equation}
Thus, different number of pulse replicas $K$ can be produced with the same aperiodic comb as shown in Fig.~\ref{fig2:grid}.

A distribution histogram characterizes the number of teeth in the comb having the same width:
\begin{equation}\label{eq-Pk}
P(d)=||\{d_m:d_m=d, m=[1,M]_{\mathbb{N}}, d\in[d_{min},d_{max}]_{\mathbb{N}}\}||,
\end{equation}
where $||S||$ denotes the size the set $S$, $d$ is the integer variable. Expression \eqref{eq-Djk} specifies $K$ interleaved subcombs, one subcomb for each pulse replica, and their union covers the whole signal processing bandwidth that is expressed as
\begin{equation}\label{eq-sum}
  \sum_{m=1}^{M}d_m = \sum_{d=d_{min}}^{d_{max}}P(d) = N.
\end{equation}
Given the distribution $P(d)$ is defined, a regular aperiodic comb with monotonically increasing tooth width can be easily constructed:
\begin{equation}\label{eq-dP}
 d_m = \min\left\{d: m\leqslant\sum_{j=d_{min}}^d P(j)\right\}.
\end{equation}
A comb vector
\begin{equation}\label{eq-Dsigma}
  D_{\sigma} = \{d_{m'}: m'=\sigma(m), m=[1,M]_{\mathbb{N}}\}
\end{equation}
is obtained by the permutation $\sigma$ applied to the initial comb vector $D_0$. Thus, a variety of different combs can be generated from the same initial comb using random permutations.

The optical frequency grid $\{\omega_n\}$ depends on the shaper configuration and its instantaneous processing bandwidth. We consider this grid fixed. Two parameters of the problem to be optimized are the comb width distribution $P(d)$ and the permutation $\sigma$ that randomizes the comb.

Numerical optimization was performed in two steps. First, the distribution minimizing average satellite amplitude over a  set of random permutations is found. Second, the permutation providing the minimum spike amplitude for the optimized distribution is found. Both steps of optimization are based on stochastic methods and do not necessarily provide the unique optimal solution of the problem (global minimum), but the results of numerical experiments were highly reproducible from test to test and robust with respect to initial conditions and parameters of the optimization algorithms.

Since the aim of the optimization was to obtain the maximum contrast of pulses to the pedestal, the cost function was based on electrical field amplitude $|E_{out}(t)|$. The diffracted  field was calculated as the FFT of modulated spectrum under assumption of Gaussian input spectrum with the effective width $\Delta\omega$,
\begin{equation}\label{eq-Eout}
  E_{out} = \mathrm{FFT}\left\{T_n \exp\left(-\frac{\omega_n^2}{2\Delta\omega^2}\right)\right\}.
\end{equation}
Exact definition of the cost function is given below. The following optimization was performed for a comb with the duty cycle $K=2$ and only one pulse replica with non-zero amplitude, $A=\{1,0\}$, because this set of parameters explicitly characterizes performance of the random comb and eliminates the effects of interference between the replicas and the satellites that can affect the algorithm performance.

\subsection{Optimization of distribution}

Searching for an appropriate distribution $P(d)$ belongs to the class of constrained optimization problems because the sum of the distribution is fixed according to~\eqref{eq-sum}. This problem can be solved using DE genetic algorithm with repair, i.e. if the optimization algorithm produces an infeasible solution, this solution is modified to satisfy the constraint~\cite{Coello02}. The parameters of the optimization problem are the values of comb tooth width $d_m$. In addition to the definition~\eqref{eq-D} and the fixed total comb span~\eqref{eq-sum}, we constrain the minimum comb tooth width $d_{min}$:
\begin{equation}\label{eq-wm}
  d_{min} \leqslant d_m
\end{equation}
The maximum comb tooth width $d_{max}$ was not constrained, but its value determined the dimension $M$ of the vector $D$
\begin{equation}\label{eq-M}
  M = \lfloor 2N/d_{max}\rfloor,
\end{equation}
where $\lfloor x\rfloor$ is the integer part of $x$. The total number of comb elements $N$ was fixed. This determined the mean value of the width as $\langle d_m\rangle\approx d_{max}/2$ for a sample of $M$ random values satisfying the constraint~\eqref{eq-sum}. The algorithm of distribution optimization was based on DE with random parent selection and binominal crossover (DE/rand/1/bin)~\cite{RoccaOliveriMassa11}.

The \emph{mutation} operator produces $D_{mut}$ from a \emph{primary parent} $D_{p1}$ and two {donor} elements of the recent population, $D_{d1}$ and $D_{d2}$
\begin{equation}\label{eq-mut}
  D_{mut} = D_{p1} + \lfloor a(D_{d1} - D_{d2})\rfloor,
\end{equation}
where $a\in(0,1]_{\mathbb{R}}$ is a random coefficient. Since $\sum(d_{d1} - d_{d2})=0$, the mutation operation conserves the sum of elements $N$. Any mutant vector that does not satisfy~\eqref{eq-wm} is immediately discarded.

Common \emph{crossover} operators used in DE algorithms randomly select vector elements either from the {secondary parent} $D_{p2}$ or from the mutant element $D_{mut}$. Binominal crossover operation is defined as:
\begin{equation}\label{eq-cross}
  D_c =  D_{p2} + B \odot (D_{mut} - D_{p2}),
\end{equation}
where $\odot$ is the Hadamard (element-wise) product of two vectors, and $B=\{b_m:m=[1,M]_{\mathbb{N}},b_m\in\{0,1\}\}$ is a random binary vector. The \emph{offspring} vector $D_c$ satisfies the constraint~\eqref{eq-wm}, but not~\eqref{eq-sum}. To fix this, we calculate a scaled offspring
\begin{equation}\label{eq-scale}
  D_s = \lfloor s D_c \rfloor, \qquad s = \frac{N}{\sum d_c}.
\end{equation}
If a scaled offspring vector does not satisfy~\eqref{eq-wm}, it is discarded and the crossover operation is immediately  repeated with different $B$. Taking the integer part of vector elements in~\eqref{eq-mut} and \eqref{eq-scale} may result in violation of sum conservation~\eqref{eq-sum}. In this case, the vectors $D_{mut}$ and $D_s$ are repaired by adding random integer numbers to some elements.

The \emph{selection} procedure is performed dynamically, i.e. the offspring vector is compared to the parents immediately after the crossover and replaces the worst of the parents in the population if it fits the selection criterion better that any of the parents.

One of specific problems in searching for an optimized width distribution $P(d)$ is that the actual pulse-to-satellites contrast depends on the permutation applied to the initial comb $D_0$ in a stochastic manner. A shuffled comb can produce sharp but narrow random spikes of $E_{out}(t)$ that sufficiently affect the performance of the DE algorithm. To cancel out this effect at the first step of optimization, we averaged the pulses over a set of random permutations before calculating the maximum pedestal intensity and performing the selection procedure. The set of 20 permutations was found to be a good compromise between smoothness of the averaged pulse shape and computation time. Thus, the optimal distribution provides the best contrast only for the statistical ensemble of permutations.

The DE algorithm was applied to a population of 100 random distributions. It was found that the optimized distribution $P(d)$ is best fitted with a power function
\begin{equation}\label{eq-Popt}
  P_{opt}(d) = \lfloor P_1 d^{\alpha} + P_0 \rfloor,\qquad \alpha<0
\end{equation}
where $\alpha$, $P_1$ and $P_0$ are the fit parameters. The values of the fit coefficients were found to be $P_1 = 2900$, $P_0=0.8$, and $\alpha = -2.2$. This distribution was further used as the optimal one.

\begin{figure}[t]
  \centering
  \includegraphics[width=0.5\columnwidth]{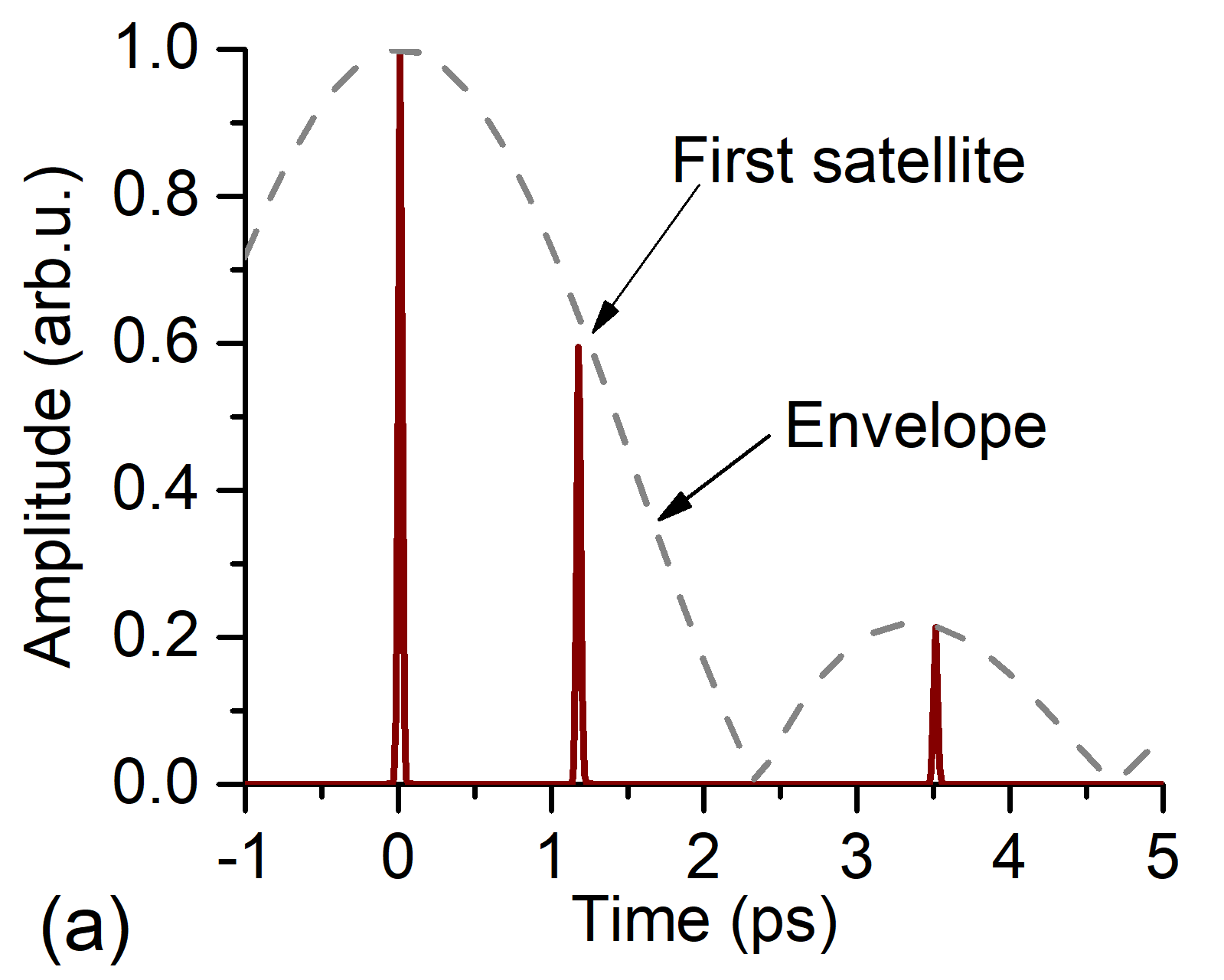}\includegraphics[width=0.5\columnwidth]{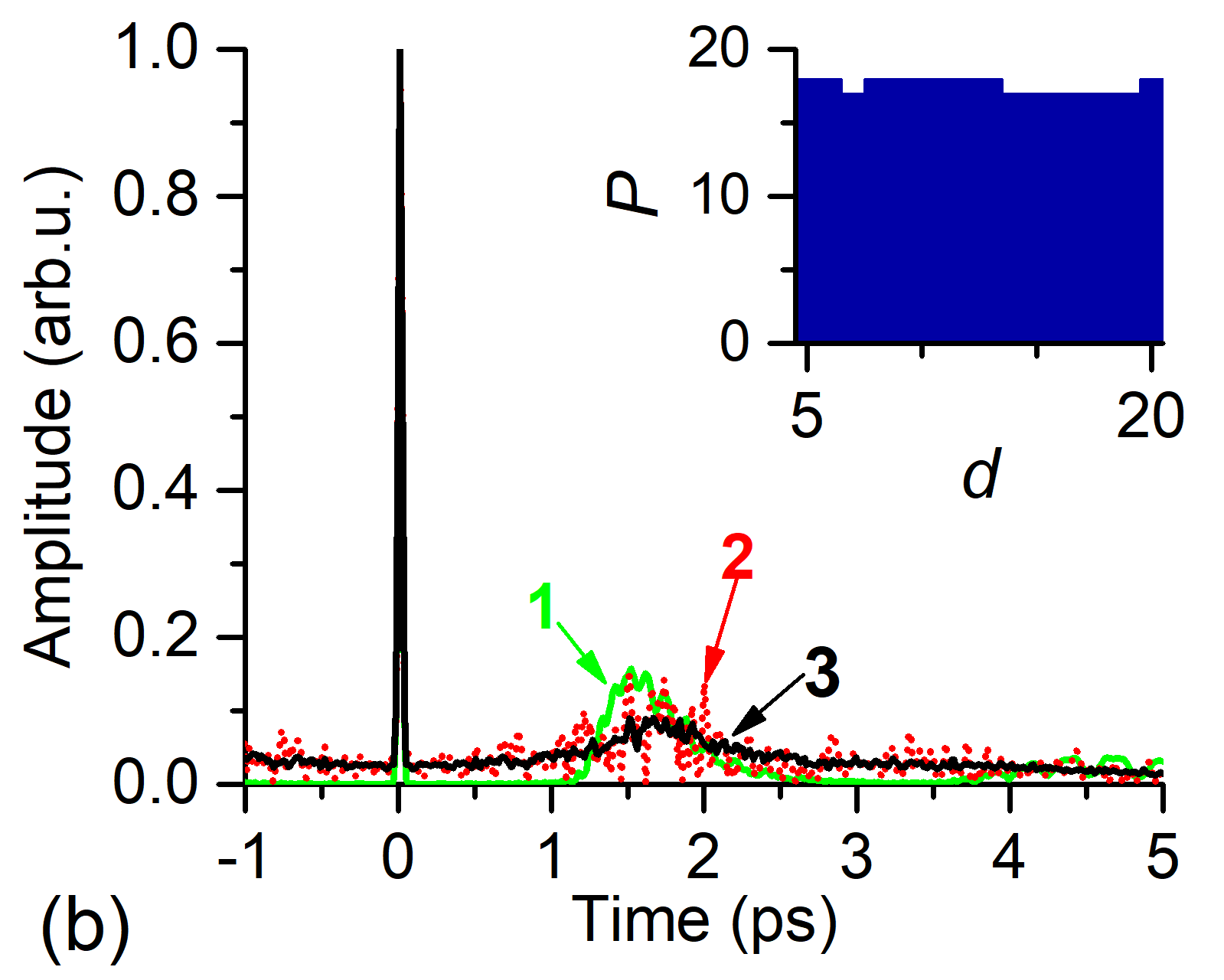}\\
  \includegraphics[width=0.5\columnwidth]{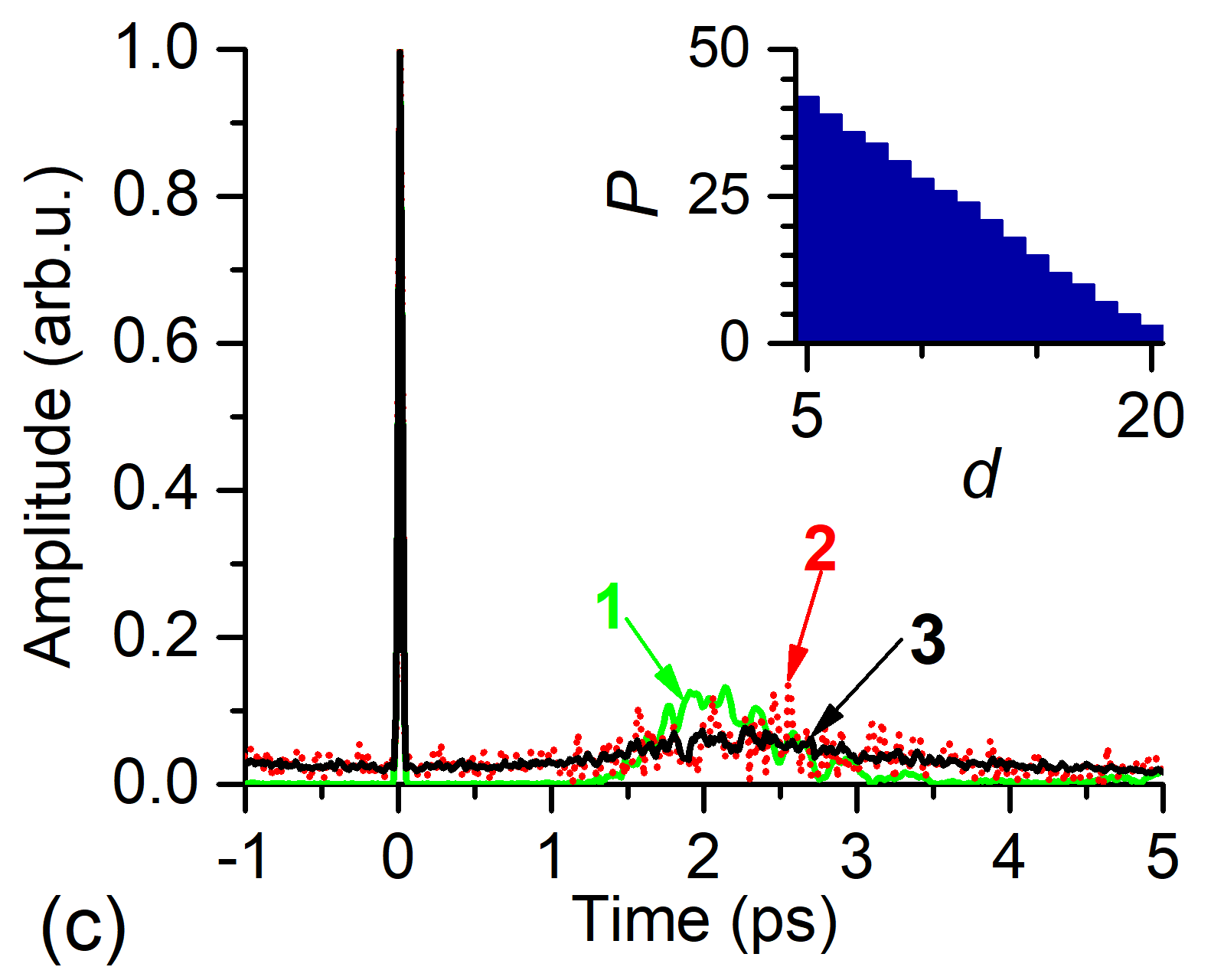}\includegraphics[width=0.5\columnwidth]{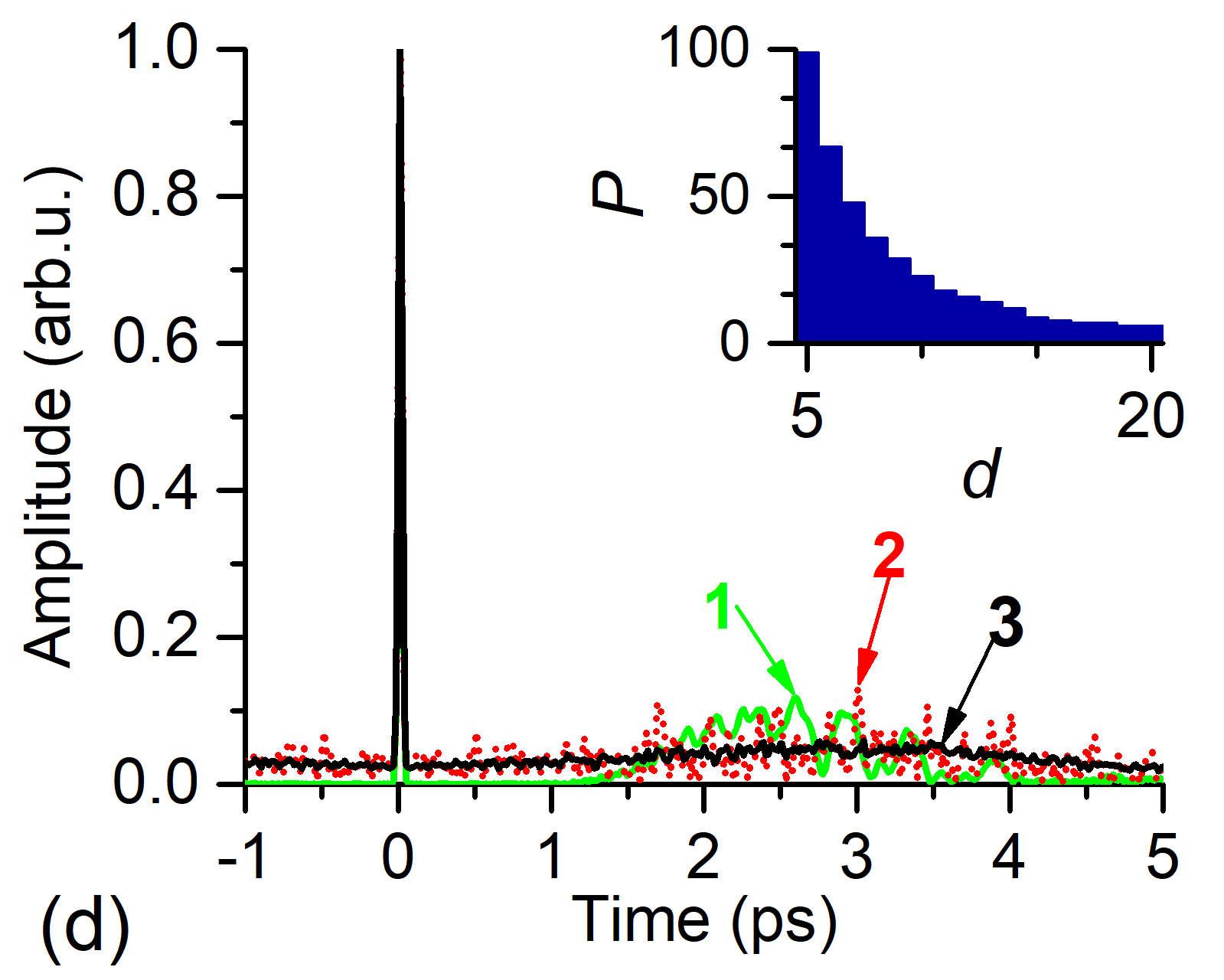}\\
  \caption{Simulated laser pulse amplitude $|E_{out}(t)|$ for a comb with duty cycle $K=2$ ($A=\{1,0\}$) demonstrates the effect of comb tooth width distribution on pulse contrast: (a) equally spaced MICS, $M=175$; (b) equal-probability distribution, $M=281$; (c) linear distribution, $M=351$; (d) optimized distribution, $M=405$. 1~--- regular comb with monotonically increasing width; 2~--- aperiodic comb with distribution $P(d)$ and a random permutation; 3~--- average of 20 random permutations used for DE optimization. Insets: distribution histogram $P(d)$.}\label{fig3:dist}
\end{figure}

The results of the optimization are shown in Fig.~\ref{fig3:dist}. We compare the satellites obtained with a standard periodic MICS with $d_0=d_{max}=20$ (Fig.~\ref{fig3:dist}a) and  three aperiodic comb distributions. All distributions have the same parameters $d_{min}=5$, $d_{max}=20$, and the sum of widths (i.e. the grid size) $N=3494$ corresponding to the parameters of the pulse shaper (see Sec.~\ref{sec-AO}). The total number of comb elements $M$ varies with the distribution. The three cases of aperiodic combs were the equal-probability distribution, $P(d)\simeq\const$, and the linear distribution, $P(d)\simeq C(d_{max}+1-d)$, and the optimized distribution $P_{opt}(d)$ defined by Eq.~\eqref{eq-Popt}. The pulse shape $|E_{out}(t)|$ in the cases (b), (c), and (d) was simulated for a regular comb with monotonically increasing width (plot 1, without a permutation), a comb obtained by its random permutation (plot 2), and an average of 20 random permutations (plot 3).

The optimized distribution~\eqref{eq-Popt} yields the maximum of the satellite amplitude lower and shifted to larger delays compared to other distributions because the comb contains more elements with smaller $d_m$ values (plot 1 in Fig.~\ref{fig3:dist}d). More important is that this distribution provides a more uniform spike amplitude after randomization of tooth order (plot 3 in Fig.~\ref{fig3:dist}d). However, large spikes in the pulse envelope can still exist when a random permutation is applied to the comb. That explains why another step of optimization is required.

\subsection{Optimization of permutation}
The second step of generating a random comb was optimization of the permutation provided the width distribution $P(d)$ is defined. The goal of permutation optimization is to eliminate the spikes of the electrical field by selecting a random permutation of the comb.

\begin{figure}[t]
  \centering
  \includegraphics[width=\columnwidth]{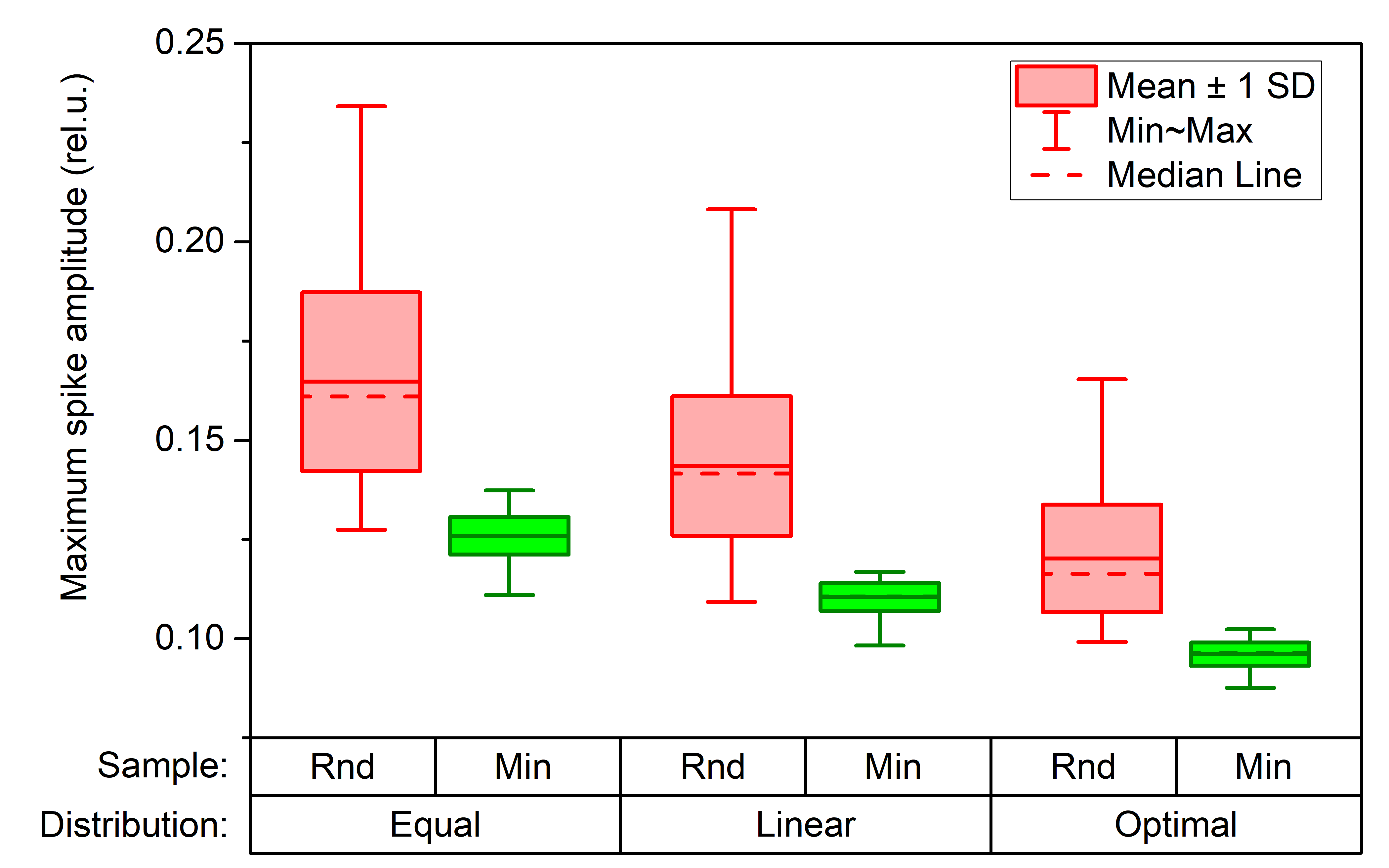}\\
  \caption{Statistics on Monte-Carlo optimization of random permutations: Rnd~--- simple random sample of 100 permutations; Min~--- a set of 10 sample minima.}\label{fig4:mc}
\end{figure}

For optimization of the random permutation we used the Monte-Carlo (MC) method that appeared to be fast and effective in rejecting the permutations with large spikes and selecting those with the smallest spikes. A simple random sample of 100 permutations was generated, and the one permutation with the minimum spike amplitude was selected from the sample. This procedure was repeated 100 times to create a set of sample minima. The permutation with the minimum spike amplitude was selected from this set. This is equivalent to selecting the permutation with the minimum cost function value $\max|E_{out}(t)|$ from a simple random sample of $10^4$ random permutations, while the whole number of different permutations is over $10^{400}$ for the optimized distribution \eqref{eq-Popt}. One should note that the search space in our case is the group of permutations but not a subset of Euclidian space, so different approaches to the optimization problem are required (see, for example,~\cite{BuchheimJunger05,SantucciBaiolettiMilani16}). We leave this problem out of scope of this work and use a simple MC method though we still mention the resulting permutation as ``optimal'' since it provides the minimum cost function over a selected simple random sample of permutations.

The statistics for different previously defined model distributions $P(d)$ is shown in Fig.~\ref{fig4:mc}. The optimized  distribution~\eqref{eq-Popt} is compared to the equal-probability distribution and the linear distribution. The results of optimization for $P_{opt}(d)$ are $\approx20$\% better compared to the equal-probability distribution and $\approx10$\% better compared to the linear distribution.

\section{Experimental results and discussion}\label{sec-exp}
\subsection{Experimental setup and methods}

\begin{figure}[t]
  \centering
  \includegraphics[width=\columnwidth]{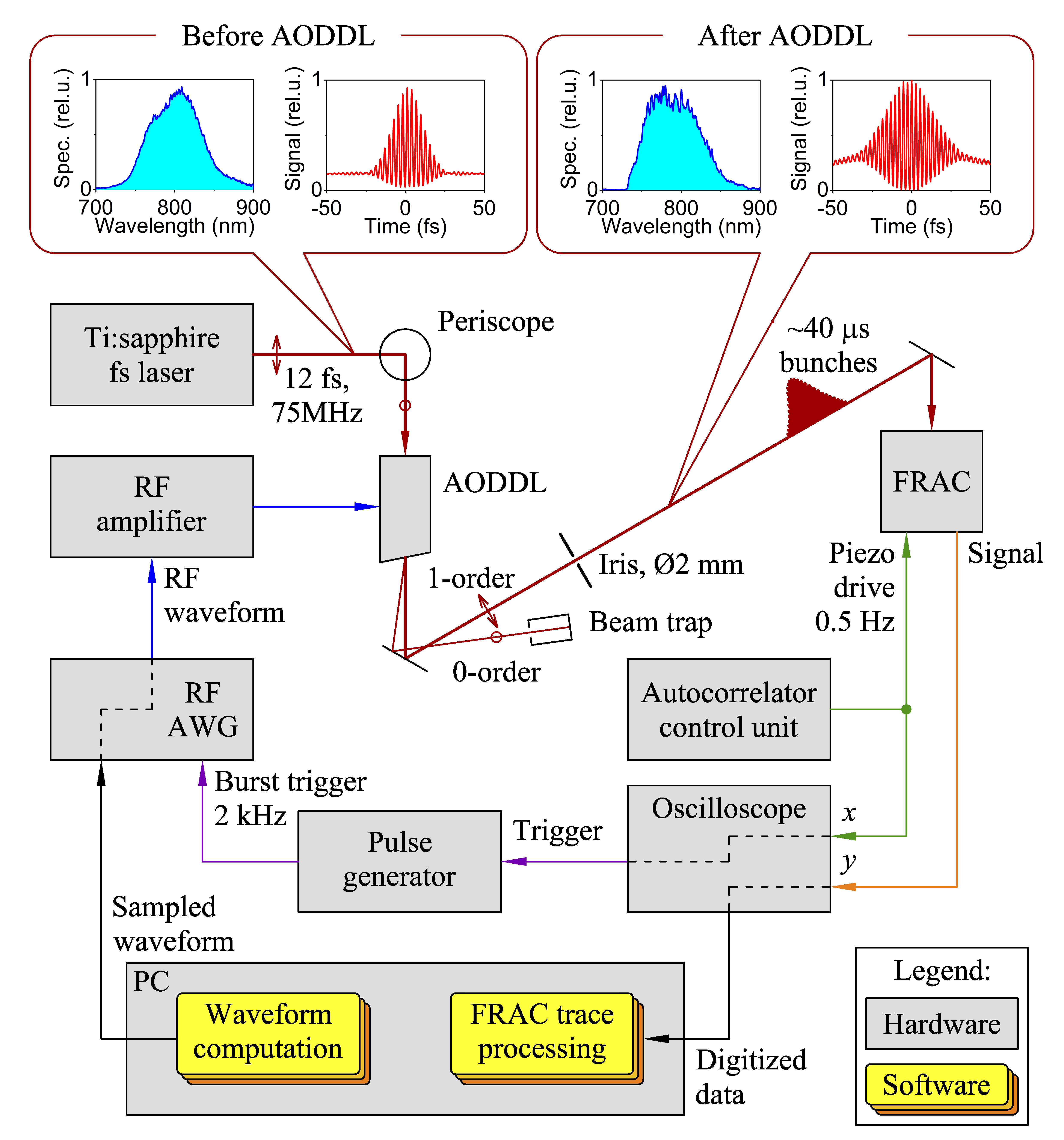}\\
  \caption{Experimental setup for RandoMICS evaluation and laser emission parameters. Insets: radiation spectrum and pulse spectrum before and after the AODDL.}\label{fig5:setup}
\end{figure}

Experimental validation of RandoMICS pulse shaping was performed with the AODDL described in Sec.~\ref{sec-AO}. The experimental setup and laser beam parameters are shown in Fig.~\ref{fig5:setup}. The spectrum and the fringe-resolved autocorrelation (FRAC) after the AODDL were measured without modulation of the spectrum, i.e. the AODDL provided flat transmission in the whole bandwidth of 150 nm ($T_n=1$ for all $n$). The RF waveforms were generated by the AWG (N8241A, Agilent Technologies) and applied to the AODDL. The pulse train of 12 fs pulses (75 nm FWHM and 135 nm FWTM bandwidth) from the master oscillator (Femtosource Synergy, Femtolasers GmbH) was directly modulated with the AODDL and measured with the scanning autocorrelator (Femtometer, Femtolasers GmbH) in FRAC measurement mode using custom signal readout with the oscilloscope (RTB2004, Rohde \& Schwarz). The AWG and the oscilloscope were synchronized with the scanning rate of the autocorrelator employing additionally a signal generator (33220A, Agilent Technologies) for triggering the AWG in a burst mode. The AODDL was operating in the negative-group-delay mode to compensate for bulk dispersion of the crystal. Diffracted laser pulses were bunched in 40 $\mu$s FWHM packets and integrated with a slow photodetector of the autocorrelator (time constant $\approx300$~$\mu$s). Since the pulses in the front and the tail of the bunch contain only a part of the whole bandwidth and dispersion compensation for them is not full, the FRAC trace at the AODDL output is wider and distorted compared to the input FRAC.

Since the electrical field amplitude $|E_{\rm out}(t)|$ of the spikes is typically below 10\% of the main peak (see Fig.~\ref{fig3:dist}), the FRAC trace amplitude is mainly determined by the interference of the main pulse with the spikes. In this case, the difference between the measured autocorrelation trace and the reference level is approximately proportional to $|E_{\rm out}(t)|$.

Prior to the experiment, the instantaneous bandwidth of the AODDL was fixed that defined the full size of the grid $N$. Then, optimization of the comb was performed in two steps as described in Sec.~\ref{sec-combs}. The result of this optimization was the randomized comb~\eqref{eq-Dsigma} with restricted tooth width $d\in[d_{min},d_{max}]_{\mathbb{N}}$. Once the optimization had been done, the comb $D_{\sigma}$ was used unchanged while different settings of the output pulse shape were applied. For each output pulse shape, the transmission $T$ was calculated as a grid function~\eqref{eq-Tn}. Finally, the RF waveform applied to the AODDL was calculated using mapping from optical to acoustic frequencies and FFT~\cite{MolchanovYushkov14,SPIE19_11210}. Thus, a randomized grid was used as a preset and synthesis of the RF waveform followed the standard and fast calculation procedure.

\begin{figure}[t]
  \centering
  \includegraphics[width=\columnwidth]{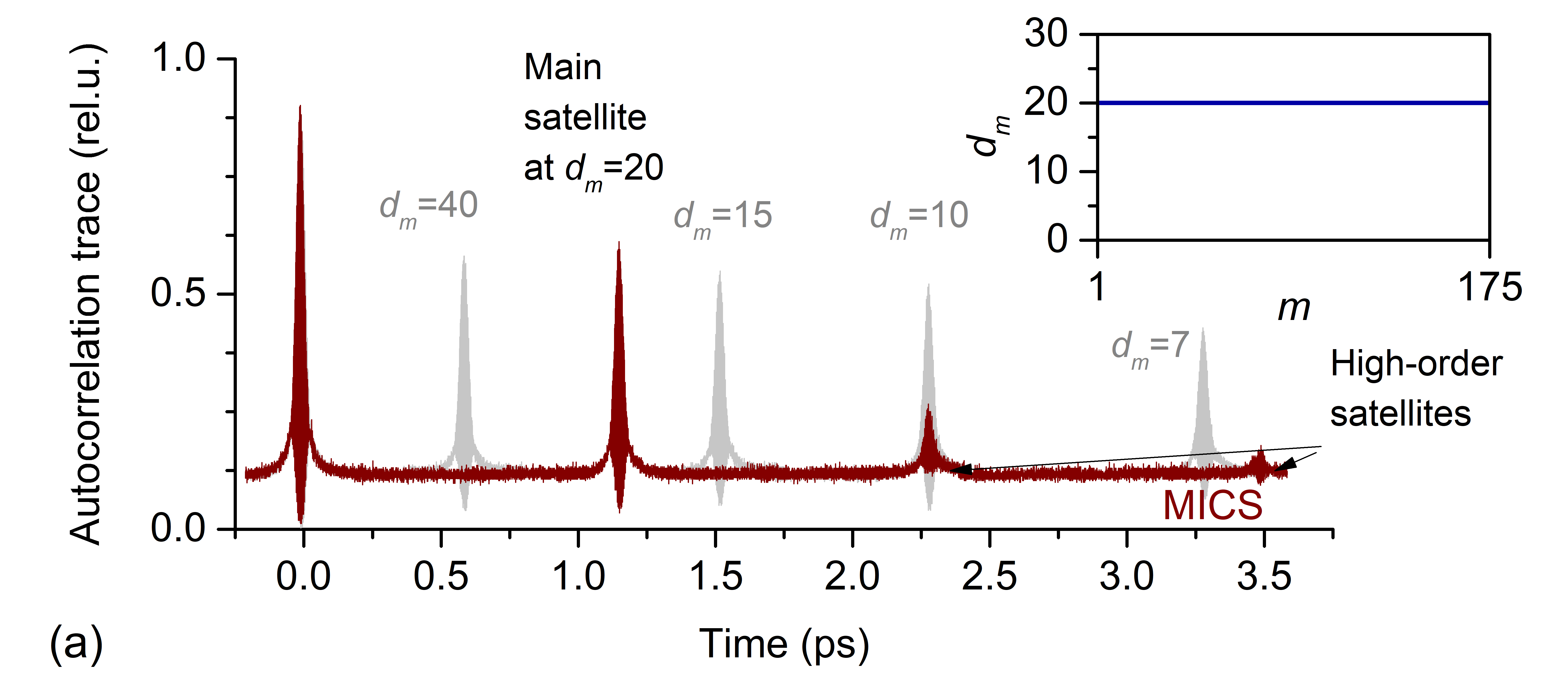}\\
  \includegraphics[width=\columnwidth]{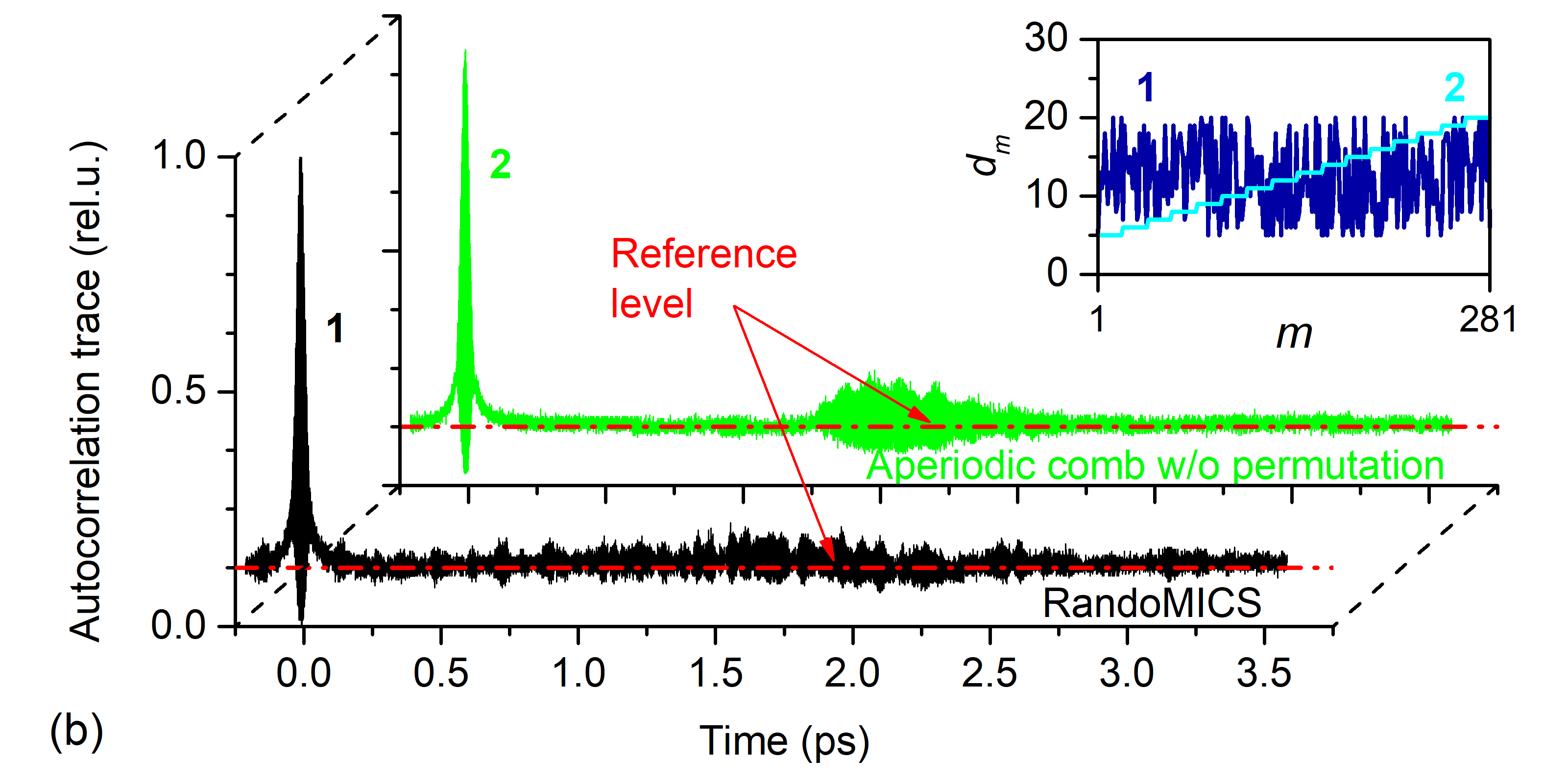}\\
  \includegraphics[width=\columnwidth]{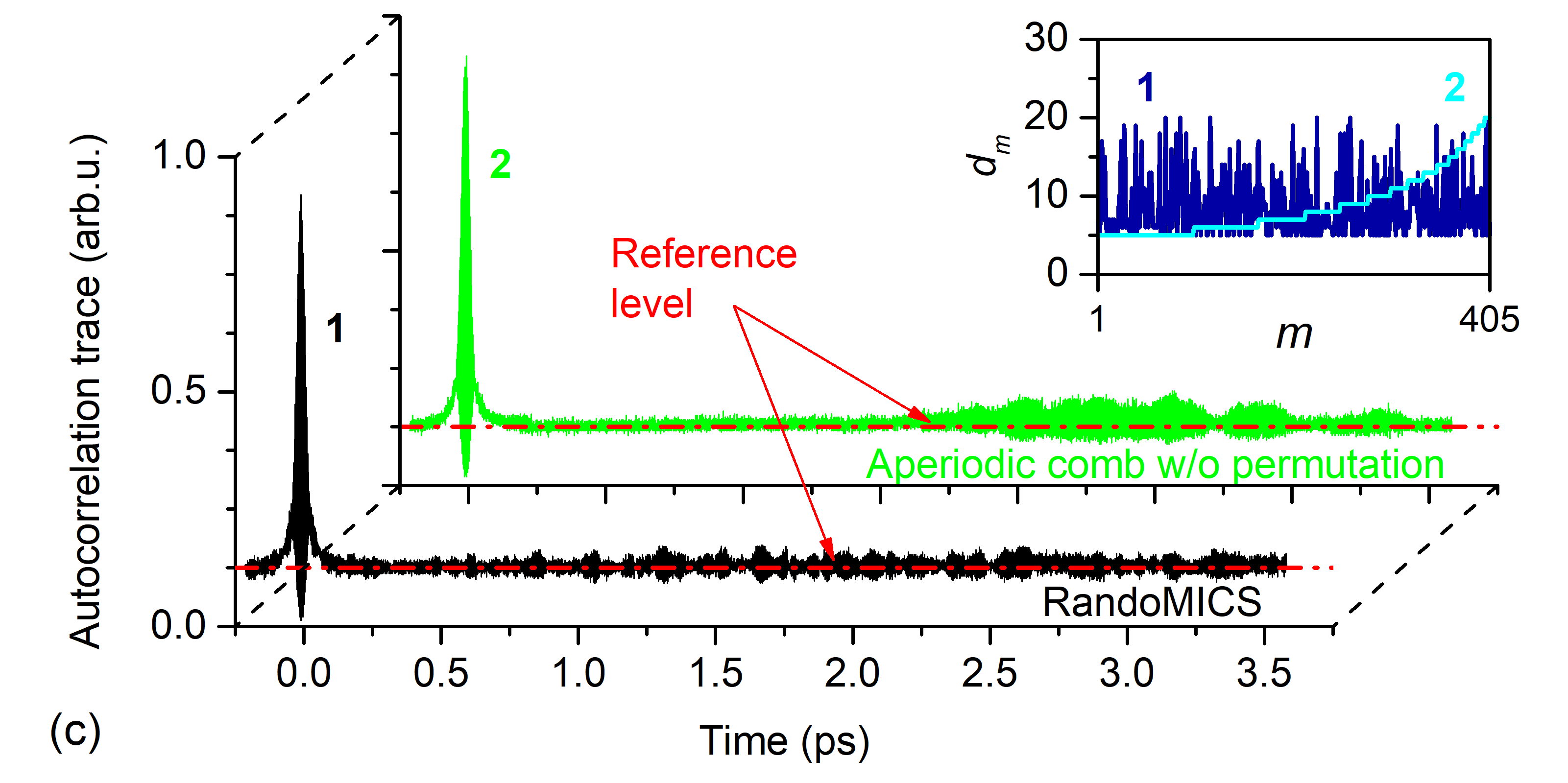}\\
  \caption{Satellite pulse suppression using randomized combs. Autocorrelation traces for the comb with duty cycle $K=2$, $A=\{1,0\}$: (a) equally spaced MICS; (b) RandoMICS with equal-probability width distribution; (c) RandoMICS with optimal width distribution. 1~--- randomized comb with optimal permutation; 2~--- regular comb with monotonically increasing width. Insets: widths of comb teeth $\{d_m\}$.}\label{fig6:acf1}
\end{figure}

\subsection{Satellite pulse  measurements}

For quantitative measurement of satellite pulse suppression we used the ability to obtain amplitude spectral modulation simultaneously with phase modulation using an AODDL. For this purpose, we generated the transmission function \eqref{eq-Tn} with the parameters $K=2$ and $A=\{1,0\}$. Thus, the spectral transmission $T_n$ was a binary pattern corresponding to the window function $W_{n1}$. That allowed measuring the amplitude of the pulse pedestal without interference between the satellites and the replicas.

Experimental autocorrelation traces are plotted in Fig.~\ref{fig6:acf1} for the regular MICS  method having a periodic comb and for the RandoMICS method with two different distributions $P(d)$. The first satellite position on the regular MICS trace (Fig.~\ref{fig7:acf2}a) was at $t\approx1.17$~ps corresponding to the comb tooth width $d_0=20$ and the frequency grid increment $\delta\omega=1.35\cdot10^{-4}$ fs$^{-1}$. Additionally, the fragments of traces with different periods of the regular comb are shown for the cases $d_m=$7, 10, 15, and 40 to demonstrate that the first satellite amplitude remains approximately the same in a wide range of comb periods. The decreasing of the satellite amplitude at smaller $d_m$ is explained by phase-to-amplitude coupling that affects overall performance of the phase-only modulation~\cite{WefersNelson95,DorrerSalin98}. Our previous estimations showed that the best performance of MICS is obtained when $d_0\gtrsim3d_{min}$ \cite{YushkovMolchanovOvchinnikovChefonov17}.

Two measurements for randomized combs with different distributions are shown in Figs.~\ref{fig6:acf1}b and \ref{fig6:acf1}c. In either case, permutation of the comb teeth makes the pedestal amplitude lower than in the regular aperiodic comb with the same width distribution (compare plots 1 and 2). An irregular structure of spikes is observed in random combs, but the optimized distribution $P_{opt}(d)$ has the maximum spike amplitude 30\% lower than the equal probability distribution (see Fig. \ref{fig6:acf1}b). The amplitude of maximum spikes in the electrical field $E_{out}(t)$ is approximately 5\% of the main peak amplitude for the case of the optimized distribution. That is lower than the first satellite obtained by the regular MICS approximately by the factor of 8.

\subsection{Pulse replication}

\begin{figure}[!t]
  \centering
  \includegraphics[width=0.5\columnwidth]{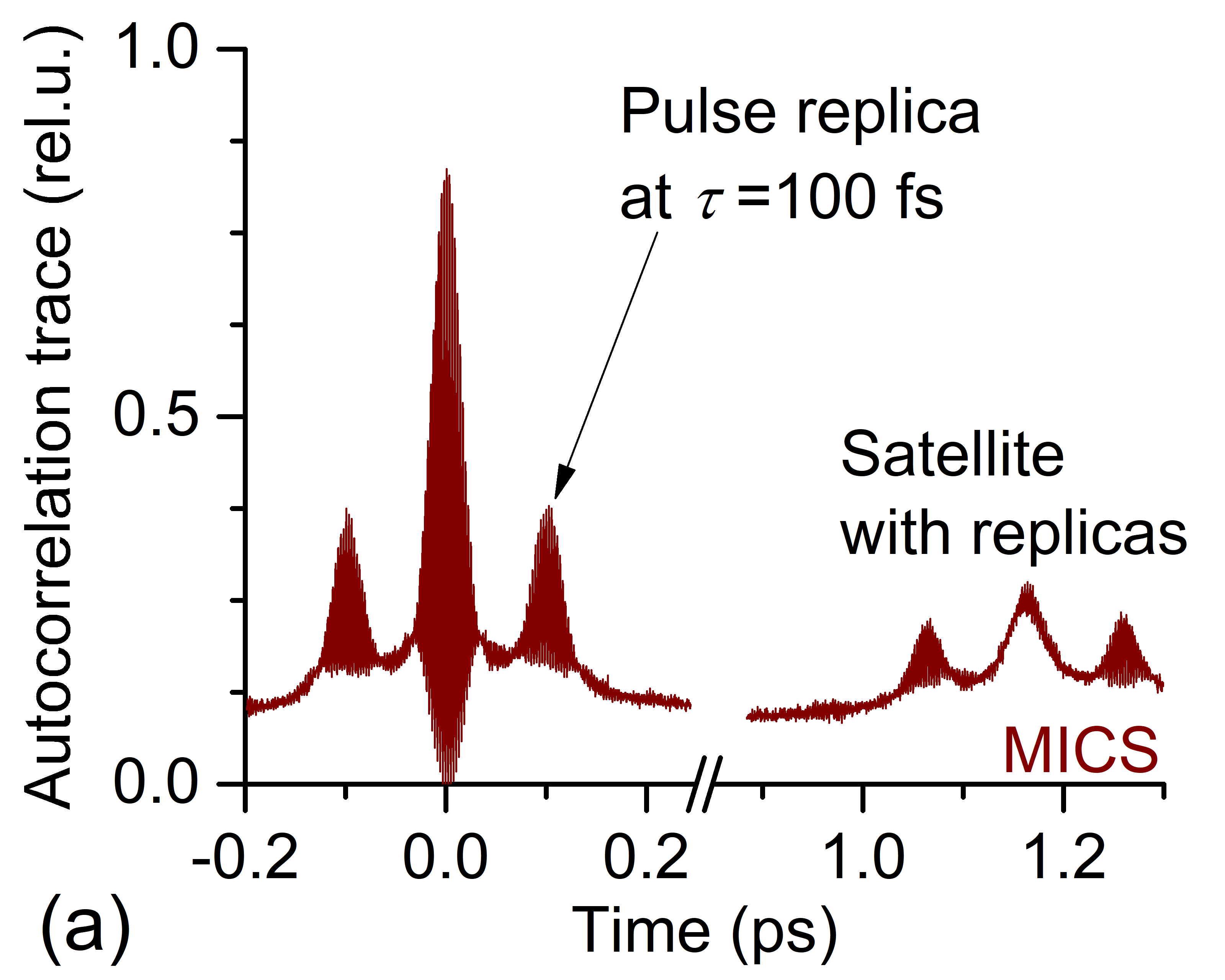}\includegraphics[width=0.5\columnwidth]{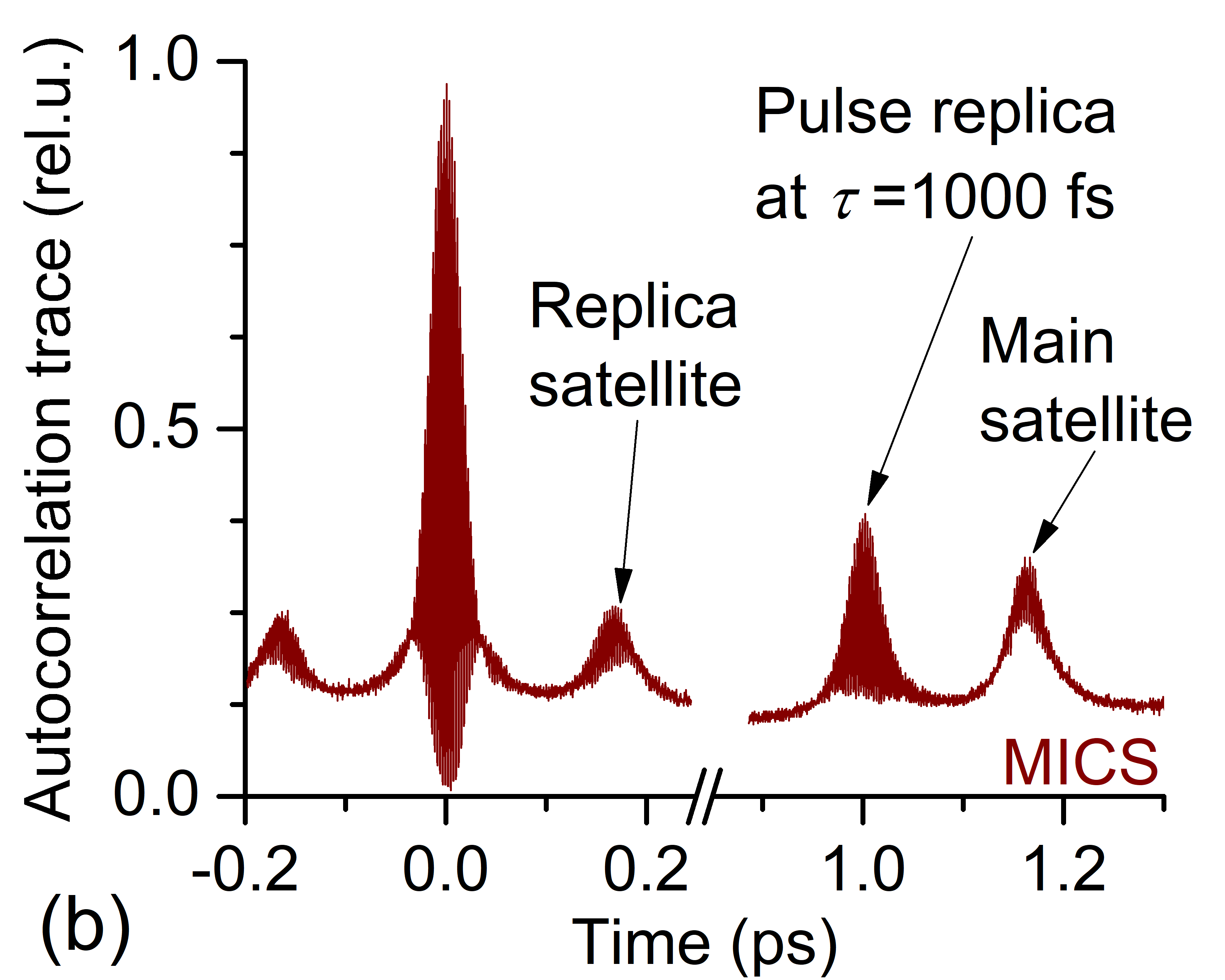}\\
  \includegraphics[width=0.5\columnwidth]{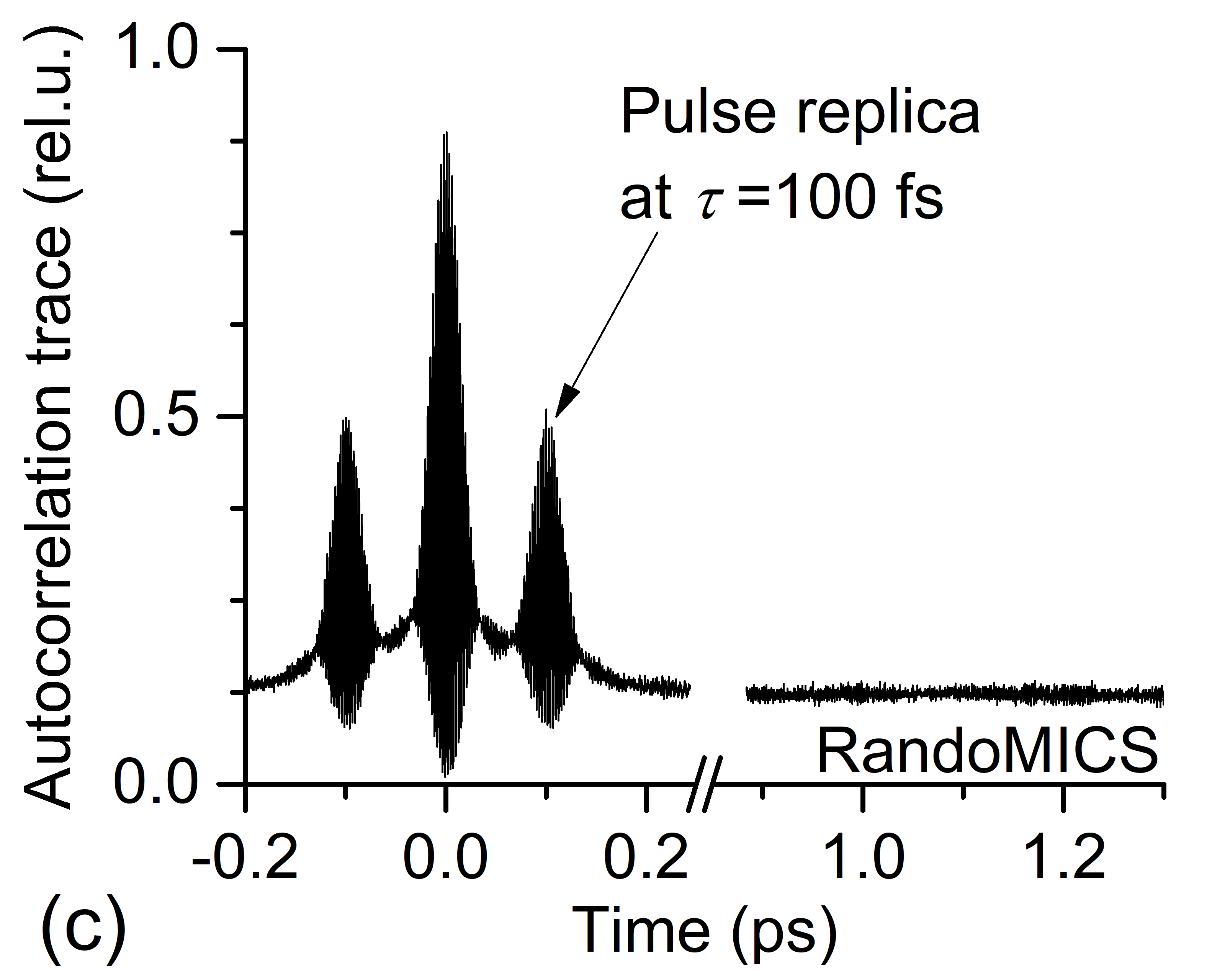}\includegraphics[width=0.5\columnwidth]{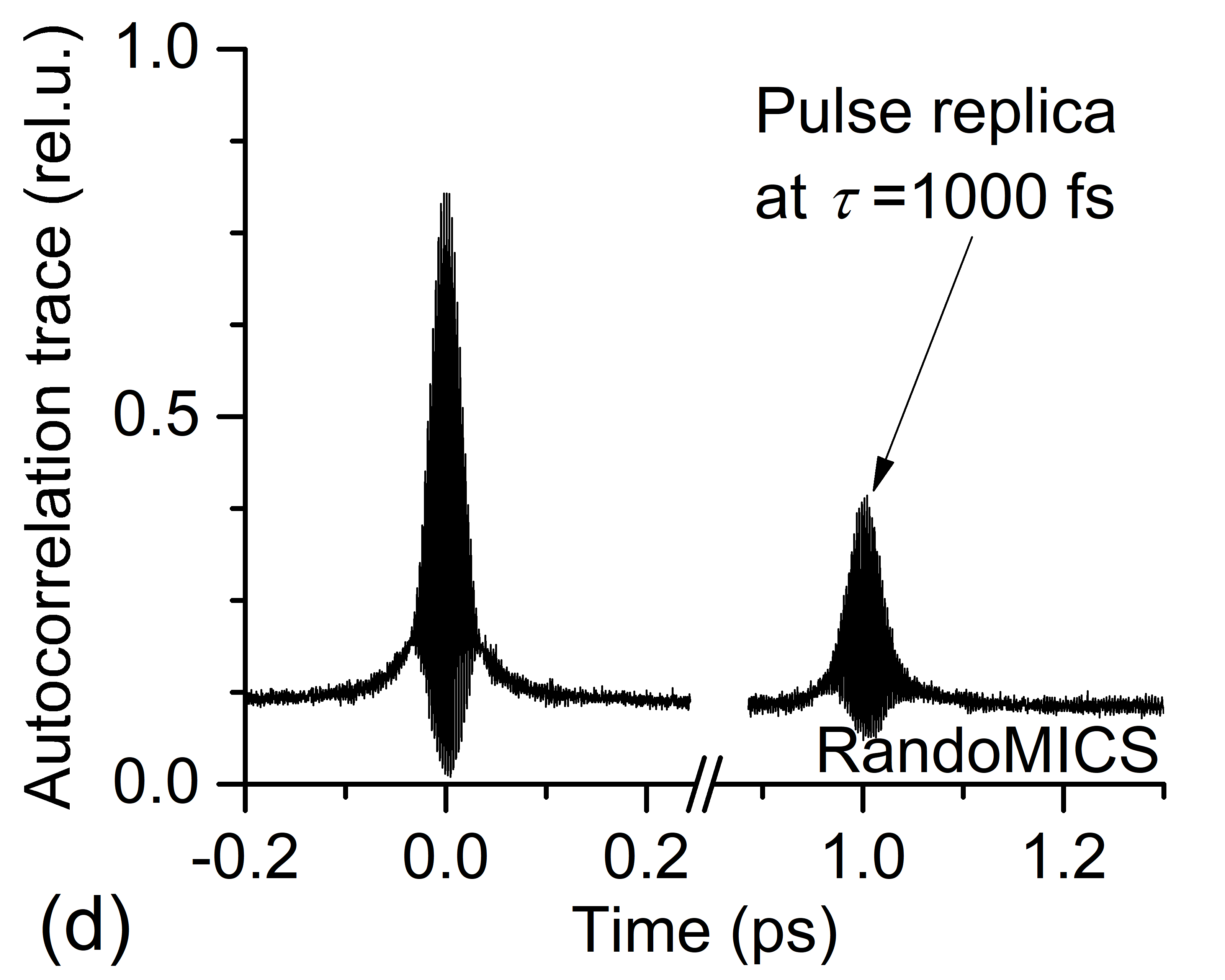}\\
  \caption{Autocorrelation traces of two replicas with delays 100 fs (left column) and 1000 fs (right column): (a,b) equally spaced MICS; (c,d) RandoMICS with optimized width distribution and permutation.}\label{fig7:acf2}
\end{figure}

Figure~\ref{fig7:acf2} illustrates the effect of comb optimization on suppression of satellite pulses for the simplest but practically important case of two pulse replicas. The experimental parameters are $K=2$, $A=\{1,1\}$, and two different delays, $\tau=100$~fs and $\tau=1$~ps. The first one corresponds to a small delay between the replicas. The second one corresponds to a large delay, when a replica is close to the satellite pulse; this case is out of scope of the regular MICS method because $\tau > \pi/(K d_0\delta\omega)$ (in this case $\max\tau=580$~fs). Regular MICS autocorrelation traces contain the main pulse and the replica, both with the satellites. The case of large delay (Fig.~\ref{fig7:acf2}b) shows the satellite close to the generated replica and comparable with it in magnitude. FRAC traces obtained with RandoMICS algorithm (Figs.~\ref{fig7:acf2}c and \ref{fig7:acf2}d) are free from satellites. This experiment explicitly demonstrates that the performance phase-only laser pulse replication technique based on interleaved orthogonal combs can be dramatically extended by making the comb aperiodic to suppress constructive interference.

\begin{figure}[t]
  \centering
  \includegraphics[width=\columnwidth]{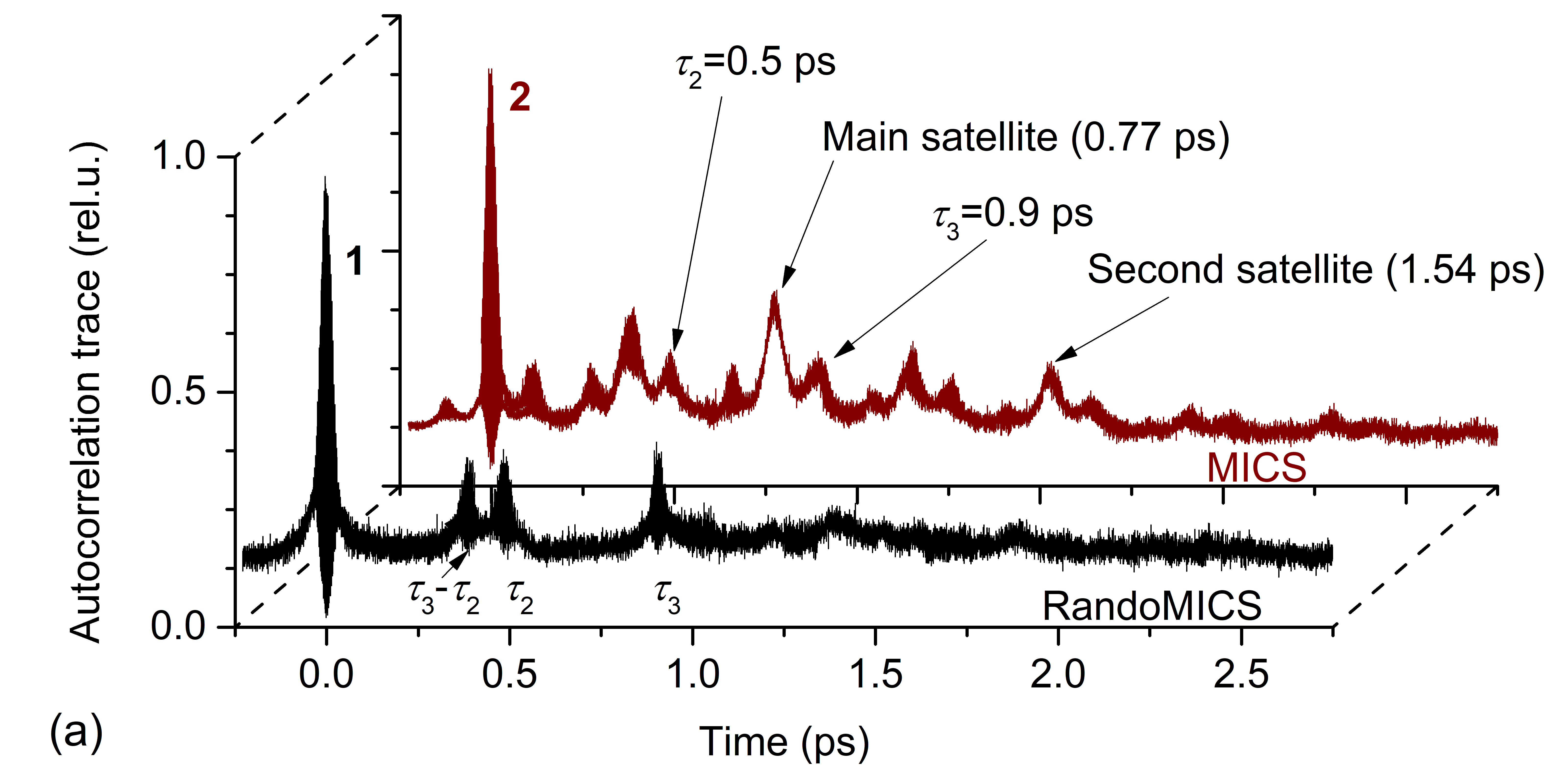}\\
  \includegraphics[width=\columnwidth]{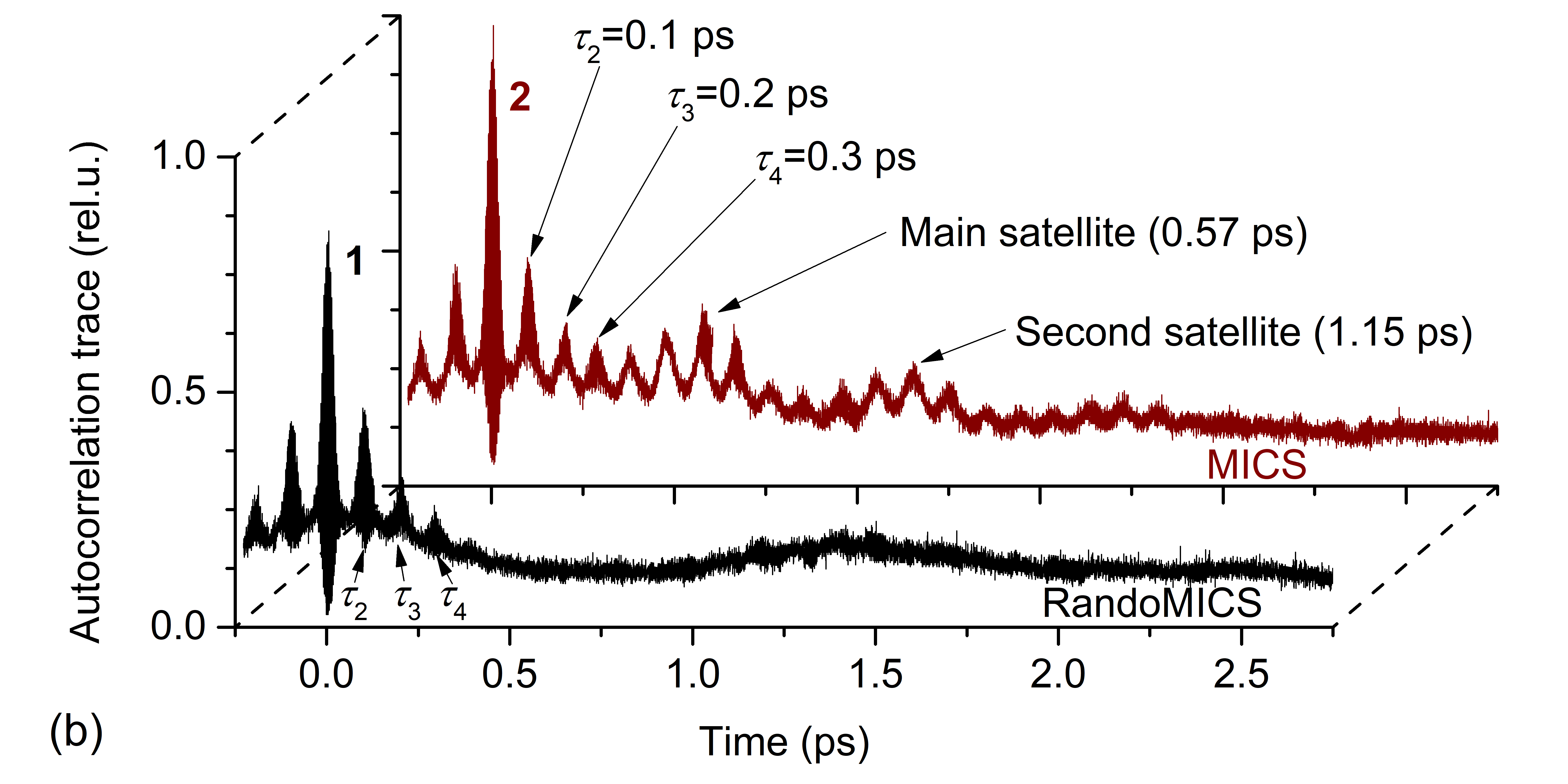}\\
  \caption{Autocorrelation traces of multiple pulse replicas: (a) uneven pulse train with $K=3$ replicas at 500 and 900 fs; (b) even pulse train of $K=4$ replicas with 200 fs spacing. 1~--- randomized comb with optimal permutation; 2~--- equally spaced comb.}\label{fig8:acf34}
\end{figure}

More complicated pulse train examples include multiple pulse replicas. Those cases are shown in Fig.~\ref{fig8:acf34} for an unevenly spaced pulse train with $K=3$ and an equally spaced pulse train with $K=4$. The delay of the first satellite pulse in a regular periodic grid is reciprocal to $K$ that can be seen comparing regular MICS traces (the maximum of the main satellite delay is 1.17 ps at $K=2$, see Figs.~\ref{fig6:acf1}a, \ref{fig7:acf2}a, and \ref{fig7:acf2}b).

The replicas in Fig.~\ref{fig8:acf34}a have the delays of $\tau_2=500$ and $\tau_3=900$ fs with respect to the main pulse, resulting in three side peaks on the FRAC trace at $\tau_2$,  $\tau_3$, and $\tau_3-\tau_2=400$ fs. The first satellite of the MICS trace is at 770 fs. There are 16 peaks in the MICS FRAC trace and they are resulting from interference of 9 pulses: three main replicas and two satellites either having two more replicas. The RandoMICS FRAC contains only three peaks and the pedestal with the maximum at 1.4 ps being $\approx2$ times smaller than the replicas.

The second example is an equally spaced pulse train in Fig.~\ref{fig8:acf34}b generating a triangular envelope of FRAC peaks with 100 fs spacing. Aliasing of the replicas and the satellites does not let to resolve them independently when regular MICS is used. On the contrary, the RandoMICS traces are free from undesired satellites that increases the usable delay range.

\section{Summary}

We proposed, elaborated, and demonstrated the advanced phase-only Randomized Multiple Independent Comb Shaping (RandoMICS) method  for ultrashort laser pulse replication based on the ability to set arbitrary programmable transmission functions of an AODDL. The described algorithm is a stochastic version of MICS with uneven tooth widths having optimized  distribution and random permutations used to reduce the amplitude of the pulse pedestal. The results include 8-fold amplitude suppression of the satellite pulses and 2-fold increase in the usable range of replica delays compared to regular MICS with a periodic comb.

The RandoMICS algorithm is implemented in two steps. The first step is optimization of the comb based on stochastic methods of differential evolution and Monte-Carlo. Once it is done, the resulting comb can be used at the second step of the algorithm for programmable pulse replication without any reconfiguration. The same random comb supports variable delays and different numbers of the replicas. This optimized irregular comb is used as a preset of the pulse shaper operating in a regular way in all other aspects. Since the algorithm does not use feedback it is not necessary to run optimization after every change of the pulse train parameters, the operation rate of the pulse shaper is not reduced compared to basic straightforward pulse shaping algorithms.

Finally, we suggest that the principle of RandoMICS can be implemented not only with AODDLs, but also with FT pulse shapers. Recent progress in SLM technology made the devices with the resolution up to 4K UHD (3840 pixel in horizontal dimension) commercially available. This number of SLM elements is quite enough to apply the described algorithm using randomized pixel binning.

\end{document}